\documentclass[11pt]{article}
\usepackage{amssymb}
\usepackage{amsmath}
\usepackage[margin=1in]{geometry}
\usepackage[parfill]{parskip}
\usepackage{physics}
\usepackage{bm}
\usepackage{comment}
\usepackage[pdftex]{graphicx}
\usepackage{sidecap}
\usepackage{caption}
\usepackage{enumerate}
\usepackage{subcaption}
\usepackage{xcolor}

\newcommand{\dup}{d_{\uparrow}}
\newcommand{\ddown}{d_{\downarrow}}
\newcommand{\dsigma}{d_{\sigma}}

\newcommand{\ddup}{d^{\dagger}_{\uparrow}}
\newcommand{\dddown}{d^{\dagger}_{\downarrow}}
\newcommand{\ddsigma}{d^{\dagger}_{\sigma}}

\newcommand{\dddsigma}{\ddsigma \dsigma}

\newcommand{\nup}{n_{\uparrow}}
\newcommand{\ndown}{n_{\downarrow}}

\newcommand{\cksigma}{c_{k\sigma}}
\newcommand{\cdksigma}{c^{\dagger}_{k\sigma}}

\newcommand{\ab}{\textit{ab initio} }

\newcommand{\tj}{\tilde{j}}
\newcommand{\tb}{\tilde{b}}
\newcommand{\tl}{\tilde{l}}
\newcommand{\tdh}{\tilde{h}}

\title{Configuration interaction approaches for solving quantum impurity models}
\author{Zuxin Jin, Wenjie Dou and Joseph E. Subotnik}

\begin{document}
	\maketitle

\section{Abstract}
We develop several configuration interaction approaches for characterizing the electronic structure of an adsorbate on a metal surface (at least in model form). 
When one can separate adsorbate from substrate, these methods can achieve a reasonable description of adsorbate on-site electron-electron correlation in the presence of a continuum of states. 
While the present paper is restricted to the Anderson impurity model, there is hope that these methods can be extended to \ab Hamiltonians, and provide insight into the structure and dynamics of molecule-metal surface interactions.

\section{Introduction}
Molecule-metal interfaces host a wide range of phenomena that are of great chemical and physical interest, including adsorption\cite{Rodriguez1992}, inelastic scattering\cite{Bunermann2015}, chemicurrents\cite{Nienhaus1999, Gergen2001}, and transport in molecular electronics\cite{Nitzan2003, Ratner2013}. 
However, quantitative predictions (and sometimes qualitative explanations)  of such processes remain a challenging task. 
Typically, two key challenges must be addressed. 
\begin{itemize}
		\item (Challenge \#1) First, the fact that a few discrete molecular levels are coupled to a continuum poses a challenge to any electronic structure calculation. 
		On the one hand, the heterogeneous nature of an interface makes the most economic solid-state tool, density functional theory (DFT), less reliable. For instance, benchmark studies show that several common functionals can lead to significantly different chemisorption energies\cite{Gautier2015, Wellendorff2015}.
		Besides, standard Kohn-Sham DFT cannot provide a good description for systems having strong multi-reference character\cite{Grafenstein2000}.
		On the other hand, the presence of a bulk metal restricts the use of accurate high-level, molecular quantum chemistry methods. Moreover, for dynamical purposes, the computational cost of an electronic structure calculation must be minimal, which disfavors many relatively accurate methods, like many-body perturbation theory (e.g. the GW approximation\cite{Hedin1965}).
		\item (Challenge \#2) Second, the presence of a continuum of electronic states around the Fermi level of a metal enables nonadiabatic effects to occur, as long as there is non-vanishing nuclear-electronic coupling. Modeling such effects adds further complexity to the problem of a molecule on a metal surface: we need appropriate methodologies for simulating nonadiabatic dynamics in the presence of a continuum of electronic states, as well as associated additional quantities (e.g. diabatic couplings or derivative couplings) from electronic structure calculations. If one can develop accurate and efficient electronic structure model of molecules on surfaces, there will be many possible applications, but this is a daunting task.
\end{itemize}

Let us now discuss the question of ground state electronic structure (challenge \#1) in more detail.
As it is difficult to benchmark electronic structure methods for realistic adsorbates on realistic substrates (due to the high computational cost), today quantum impurity models mimicking adsorbates on metal surfaces usually serve as test beds for numerical solvers.
One famous example is the Newns model of hydrogen chemisorption\cite{Newns1969}, where an Anderson impurity model (AIM) with on-site repulsion $U\neq 0$ can be approximately solved with the Hartree-Fock approximation.
While reasonable chemisorption energies for several metals can be obtained, the total charge on the  hydrogen is systematically overestimated.
To go beyond Hartree-Fock, there are now several powerful numerical methods available, e.g. quantum Monte Carlo (QMC)\cite{Gull2011}, numerical renormalization group (NRG)\cite{Bulla1997,Bulla2008}, and exact diagonalization (ED)\cite{Caffarel1994}. Although QMC and NRG can in principle give accurate solutions to the AIM, these methods demand a lot of computational effort. Obviously, the size of an ED calculation scales exponentially with the number of bath orbitals, and approximations must be made in order to reduce the error introduced by bath discretization (for instance, a truncation based on configuration interaction\cite{Zgid2012}).


Now, let us turn to the question of generating multiple electronic states (challenge \#2).
In the past few decades, nonadiabatic effects have been identified to play an important role in many molecular interfacial processes\cite{Morin1992, Wodtke2000, Bunermann2015}.
To model such effects, it is necessary to take multiple potential energy surfaces (PESs) into consideration.
There are, however, a few complexities in this aspect.

First, if we are interested only in molecular nonadiabatic behavior, we might expect that we would need a description with just a few electronic degrees of freedom (DoFs) corresponding to the relevant molecular diabatic states. Indeed, for molecular systems, such states can be found through several diabatization approaches\cite{Baer1975, Baer1980,Pacher1988,Pacher1991,Ruedenberg1993,Ruedenberg1997,Truhlar2001,Truhlar2002,Truhlar2003,Subotnik2008,Subotnik2009}. Nevertheless, in the presence of a metal (or semiconductor), these discrete levels will be extended between molecule and metal, and involve a continuum of crossing points, where a molecular picture is not directly identifiable.
This scenario is closely related to low-energy electron-molecular scattering\cite{Feshbach1962, OMalley1967, Domcke1983, Domcke1991}, where one typically uses a projection-operator formalism to select channels of interest.
In a \textit{single}-electron picture, Kondov et al have generalized this concept to adsorbate-substrate systems and successfully performed a diabatization at a dye-semiconductor interface\cite{Kondov2007}.
More generally, block-diagonalization of the Fock matrix is standard in transport calculations\cite{Brandbyge2002}
In a \textit{many}-electron framework (with electron-electron interactions), however, such diabatization is still a very challenging task and is limited to small cluster substrates.

Second, for closed molecular systems, DFT/TDDFT is known to give incorrect predictions of the dimensionality of conical intersections between the singlet ground PES ($S0$) and the lowest singlet excited PES ($S1$)\cite{Levine2006, Gozem2014}.
Nevertheless, DFT remains the first choice for electronic structure in many scenarios because of its wide applicability, mild scalability, and reasonable balance between cost and accuracy make.
For this reason, various approaches have been proposed to (more or less) address the issue of $S0-S1$ crossing in DFT. For instance, if one can expect certain charge character, constrained-DFT is a powerful tool to generate a meaningful diabatic representation\cite{Kaduk2012} with possible application to conical intersections\cite{Kaduk2010}.
More generally, for small systems, one may resort to multiconfiguration\cite{Grafenstein2005} or multireference \cite{Grimme1999, Levine2006} DFT methods.
Recently, based on studies of double excitation states in TDDFT\cite{Maitra2004, Cave2004, Laikov2007}, 
Teh et al suggested that, by merely adding one selected double excitation to HF/CIS or DFT/TDDFT (so called CIS-1D or TDDFT-1D), one can recover the correct $S_0-S_1$ conical intersection topology with reasonable energetic accuracy\cite{Teh2019}.
Thus, to date, there has been some progress improving DFT to allow for static correlation in the \textit{gas phase}. 
Nevertheless, for the most part, these DFT methods have not been applied to molecules on \textit{metal} surfaces, where charge transfer is possible.
In general, if one were to study a molecule on a metal surface with a well-developed embedding theory, e.g., DMET\cite{Knizia2012, Wouters2016}, one would generally be interested in wave-function methods, especially in the case that electron correlation is largest between molecular electrons, not between molecular and metal electrons. More often than that, one freezes the electrons in the bulk metal\cite{Govind1998,Govind1999}.
This will be discussed again below.

From the discussion above, it is clear that, in order to model adsorbate dynamics on a metal or insulating surface, many challenges and opportunities remain. At bottom, one requires a robust electronic structure approach that can generate a finite set of electronic states in the presence of a continuum of states and nontrivial electron-electron repulsion.
With this goal in mind, in the present work, we will extend the idea of CIS-1D\cite{Teh2019} mentioned above to study charge character in a molecular-metal system.
Specifically, we will investigate the electronic structure of the Anderson impurity model from a configuration interaction approach, and compare the ground state molecular charge with the exact answer from NRG.
Subsequently, based on these approaches, a projection-based diabaziation is proposed to generate a diabatic picture for the system's many-electron states.

The paper is organized as follows. In Sec. \ref{sec:method}, we introduce our configuration interaction-based electronic structure method and projection-based diabatization. The results of these methods applied to Anderson-Holstein model are presented in Sec. \ref{sec:result}. A discussion of these methods are given in Sec. \ref{sec:discussion}. We conclude in Sec. \ref{sec:conclusion} with an outlook for future dynamical applications.

Regarding notations below, $i,j,k, \ldots$ label canonical Hartree-Fock (mean-field) occupied orbitals, $a,b,c,\ldots$ label canonical virtual orbitals. A tilde over an occupied (virtual) orbital means that it is a linear combination of canonical occupied (virtual) orbital. A bar above an orbital represents spin-down, and orbitals without bars are assumed spin-up.

\section{Methods}\label{sec:method}
For this paper, we will work with the Anderson impurity model (AIM), 
\begin{align}\label{eq:H:AH}
H = E(x) \sum_{\sigma} \dddsigma + U \nup \ndown  + 
\sum_{k\sigma} \left( V_k \cdksigma \dsigma  + V_k^*\ddsigma  \cksigma \right) + 
\sum_{k\sigma} \epsilon_k \cdksigma \cksigma
\end{align}
Here, $\dsigma$ corresponds to an impurity orbital with spin $\sigma$ and orbital energy $E(x)$ ($x$ represents a nuclear coordinate).
$\nup \equiv \ddup \dup $ $(\ndown\equiv\dddown\ddown)$ is the particle number operator.
$U$ is the impurity on-site repulsion.
$\cksigma$ corresponds to the bath state labeled by $k$ with spin $\sigma$ of energy $\epsilon_k$. 
$V_k$ is the system-bath coupling amplitude.
The bath spectrum $\{\epsilon_k\}$ forms a quasi-continuum. We will now present a host of approaches for solving the AIM.

\subsection{Restricted Mean-Field}
We begin with the simplest approximation, mean-field theory, whose behavior is well-known.
A crude ground state for Eq. \ref{eq:H:AH} can be obtained from the (restricted) Hartree-Fock approximation. If we assume the ground state can be expressed by a closed-shell single Slater determinant,
\begin{align}\label{eq:gs:mf}
\ket{\Psi_0} = \ket{i\bar{i} j \bar{j} \ldots }
\end{align}
the variational principle yields the (spinless) self-consistent equation:
\begin{align}\label{eq:scf}
(h + Un_0 \dyad{d}{d}) \ket{i} = \lambda_i \ket{i}
\end{align}
where $h$ is the one-body part of the Hamiltonian, and
\begin{align}\label{eq:pop}
n_0 \equiv \sum_{j} \abs{\ip{d}{j}}^2
\end{align}
is the population of the impurity orbital $\ket{d}$ for one spin ($\uparrow$ or $\downarrow$).

In the limit of $U \rightarrow 0$, the model is essentially non-interacting.
In the wide band limit, the orbital population takes the simple form\cite{Bruus2004}
\begin{align}\label{}
\ev{\nup} = \ev{\ndown} = \frac{1}{\pi}\int \dd\epsilon  \frac{\Gamma/2}{(\epsilon - E(x))^2 + (\Gamma/2)^2} f(\epsilon,\mu)
\end{align}
here, $\mu$ is the chemical potential, $\Gamma \equiv 2\pi \sum_{k} \abs{V_k}^2 \delta(\epsilon-\epsilon_k)$ is the hybridization, and $f$ is the Fermi function.

The ansatz of a single determinant can qualitatively break down in the case of $U\gg \Gamma$. Note that Eq. \ref{eq:gs:mf} has an inherent mean-field behavior:
\begin{align}\label{eq:mfprop}
\ev{\nup \ndown}{\Psi_0} = \ev{\nup}{\Psi_0} \ev{\ndown}{\Psi_0}
\end{align}
While this relation should hold when $(\mu-E(x)-U)/\Gamma \gg 1$, where $\ev{\nup} = \ev{\ndown} \approx \ev{\nup\ndown} \approx 1$, or $(E(x)- \mu)/\Gamma \gg 1$, where $\ev{\nup} \approx \ev{\ndown} \approx \ev{\nup\ndown} \approx 0$, it becomes problematic when $\mu-U \lesssim E(x) \lesssim \mu$.
On the one hand, the non-interacting part of the Hamiltonian would prefer the system to be significantly occupied; on the other hand, the large repulsion $U$ would prevent such occupation.

A general many-electron state does not necessarily suffer from this issue; significant population, zero net spin ($\ev{\nup} = \ev{\ndown}$) and small on-site repulsion can happen simultaneously. For example,
\begin{align}\label{eq:tworef}
\ket{\psi} \approx \ldots + \ket{\ldots, c_k  \bar{d} } + \ket{\ldots, \bar{c}_k  d} + \ldots
\end{align}
In other words, a superposition of singly-occupied determinants should be energetically preferred over a restricted mean-field ground state in the correlated regime.

The artifact of a restricted mean-field ground state will lead to a qualitatively wrong impurity population as a function of nuclear coordinate: regardless of the magnitude of $U$, the estimate of the impurity population changes smoothly from 0 to 1 as $E(x)$ moves from above to below the chemical potential, as one can readily observe from the mean-field spectral function\cite{Bruus2004} (note that $\ev{\nup} = \ev{\ndown}$)
\begin{align}\label{}
A(\epsilon,\uparrow)= \frac{1}{\pi} \frac{\Gamma/2}{(\epsilon-E(x)-U\ev{\ndown})^2 + (\Gamma/2)^2}
\end{align}
By contrast, given $U \gg \Gamma$, the exact impurity population should change drastically near two positions, $x_1$ and $x_2$, where $E(x_1) = \mu$ and $E(x_2) + U = \mu$, and display a population plateau (Mott plateau) in between. 

Note that the relationship between the ground-state impurity population and nuclear coordinate carries charge transfer information, which is closely related to nuclear dynamics. In the case of large $U$ and not large $\Gamma$, nuclear dynamics may well exhibit strongly non-adiabatic behavior near $x_1$ and $x_2$, and be almost adiabatic elsewhere. A mean-field description, however, would incorrectly predict a mildly non-adiabatic dynamics over a wide range between $x_1$ and $x_2$. 

\subsection{Configuration Interaction (CI)}
The discussion above (and Eq. \ref{eq:tworef}) suggests that single excitations should be crucial for optimizing the ground state. However, if one runs a variational calculation in the enlarged space $\{\ket{\Psi_0}, \ket{\Psi_i^a} \}$, the global ground state will remain due to the Brillouin theorem:
\begin{align}\label{}
\mel{\Psi_0}{H}{\Psi_i^a}=F_{ia}= 0
\end{align}
Of course, once double excitations are involved, the ground state and single excitations can be indirectly coupled and the ground state can change. 
For our purpose, however, a practical question arises: if we include double excitations in a CI calculation that model bath orbitals explicitly, will not the number of configuration states become formidably large even at the level of CISD?
While a CIS calculation uses a basis of $N_{occ} N_{vir}$ states, a full CISD calculation would involve $(N_{occ} N_{vir})^2$ doubles, which would be impossible in practice. 
After all, formally, we are modeling a true bath when $N_{occ} N_{vir}$ are infinite. 
Thus, even after discretizing the bath, the set of configuration states must be further tailored, and one would like to apply selective-CI methods to the AIM.

Below, several configuration interaction schemes are presented. A complete list can be found in Table \ref{table:CI}.
\begin{center}
	\begin{tabular}{ c |  c | c }
		Method & Configuration State Basis & Basis Size \\ \hline
		Three-state (CAS(2,2)) & $\ket{ \Psi_0  } , \ket{ \Psi_{\tdh}^{\tl}  }_s , \ket{ \Psi_{\tdh \bar{\tdh}}^{\tl\bar{\tl}  }  } $  & 3\\ 
		CIS-1D & $\ket{\Psi_0}, \{ \ket{\Psi_i^a}_s  \}, \ket{\Psi_{\tdh\bar{\tdh}}^{\tl\bar{\tl}}  }$ & $N_{occ} N_{vir}+2$ \\ 
		CIS-ND & $\ket{\Psi_0}, \{ \ket{\Psi_i^a}_s  \}, \{ \ket{\Psi_{\tdh \tj}^{\tl \tb} }_+  \} $ & $2 N_{occ} N_{vir}+1$\\ 
		MRCIS &$\ket{\Psi_0}, \{ \ket{\Psi_i^a}_s  \}, \{ \ket{\Psi_{\tdh\tj}^{\tl\tb} }_+  \} , \{\ket{ \Psi_{\tdh \bar{\tdh}\tj}^{\tl\bar{\tl}\tb  }  }_s \}$ &  $3N_{occ}N_{vir} - (N_{occ}+N_{vir}) + 2$ \\
	\end{tabular}
\captionof{table}{ A set of four different selective CI calculations for solving the AIM (Eq. \ref{eq:H:AH}). We list here both the basis functions as well as the size of the selective CI space. Orbitals with a tilde are rotated canonical orbitals. $\tdh$ and $\tl$ are active orbitals defined in Eqs \ref{eq:proj:h}-\ref{eq:proj:l}.  A subscript ``s'' means the states are singlets, e.g.. $\ket{\Psi_i^a}_s \equiv ( \ket{\Psi_i^a} + \ket{\Psi_{\bar{i}}^{\bar{a}}  } )/\sqrt{2}$. {A subscript ``+'' denote states that are linear combinations of opposite-spin determinants but not spin-adpative: $\ket{\Psi_{\tdh \tj}^{\tl \tb} }_+ \equiv \frac{1}{\sqrt{2}} \left(  \ket{\Psi_{\tdh\bar{\tj}}^{\tl\bar{\tb}}} + \ket{\Psi_{\bar{\tdh}\tj}^{\bar{\tl}\tb}} \right)$}  }
\label{table:CI}
\end{center}

\subsubsection{Three-state (CAS(2,2) with DMET Active Space)}
In molecular systems, the simplest approach to capture multi-reference character is the complete active space (CAS) method. In a CAS(M,N) calculation, one first identifies N orbitals as active orbitals, and the ground state is diagonalized in the space spanned by all configuration state functions with M valence electrons populated in N active orbitals. 
This approach can address the wrong topology of HF/CIS conical intersections and model bond making and bond breaking processes.
While for molecular systems, active orbitals often be chosen reasonably from the valence orbitals near the Fermi level, for our purposes, we will require a different strategy for picking active orbitals, as there is essentially a continuum of states near the Fermi level.

In the present work, the active orbitals are determined by a projection scheme,
\begin{align}
\ket{\tdh} &\equiv N_h \sum_{j}^{occ} \ket{j} \ip{j}{d} \label{eq:proj:h}\\
\ket{\tl} &\equiv N_l \sum_{b}^{vir} \ket{b} \ip{b}{d} \label{eq:proj:l}
\end{align}
where $\ket{d}$ is the impurity orbital and $N_h$ and $N_l$ are normalization constants. 
Note that $\ip{\tilde{l}}{\tilde{h}}=0$. For the AIM, the active orbitals in Eqs. \ref{eq:proj:h}-\ref{eq:proj:l} are essentially the same as the ones used by density matrix embedding theory (DMET)\cite{Knizia2012, Wouters2016}. 
Note that $\ket{d} = \ket{\tilde{h}}/N_h + \ket{\tilde{l}}/N_l$, and so the subspace spanned by $\{\ket{\tilde{h}} ,\ket{\tilde{l}}  \}$ is the same as the one spanned by $\{\ket{\tilde{h}} ,\ket{d}  \}$, which is the DMET active space. 
In the context of DMET, $\ket{\tilde{h}}$ is the Schmidt impurity orbital corresponding to $\ket{d}$, and the Schmidt bath orbital is proportional to $(1-\dyad{d}{d}) \ket{\tilde{h}} = \ket{\tilde{h}} - N_h\ev{n} \ket{d}$.
Here, our specific choice of the two orbitals in the active space bears a charge-transfer meaning:
if on-site energy $E(x)$ is far below $\mu_F$, $\ket{\tilde{h}} \approx \ket{d}$ and $\ket{\tilde{l}}$ is some orbital localized in bath; if $E(x)$ is far above $\mu_F$, $\ket{\tilde{l}} \approx \ket{d}$ and $\ket{\tilde{h}}$ is localized in bath; if $E(x) \approx \mu_F$, there is $\sum_{j}^{occ} \abs{\ip{j}{d}}^2 \approx \sum_{b}^{vir} \abs{\ip{b}{d}}^2 $, and $\ket{\tilde{h}} - \ket{\tilde{l}}$ should be localized to the bath and form the natural bath complement to the impurity orbital $\ket{d}$.

With the active orbitals give by Eqs. \ref{eq:proj:h}-\ref{eq:proj:l}, one can perform a ground state optimization in the sense of a CAS(2,2) calculation, i.e., the optimized ground state
\begin{align}\label{}
	\ket{ \Psi_g } = c_0 \ket{\Psi_0} + c_1 \ket{ \Psi_{\tdh}^{\tl}  }_s + c_2 \ket{ \Psi_{\tdh \bar{\tdh}}^{\tl\bar{\tl}  }  }
\end{align}
will be obtained through diagonalizing the Hamiltonian the subspace $\{\ket{ \Psi_0  } , \ket{ \Psi_{\tdh}^{\tl}  }_s , \ket{ \Psi_{\tdh \bar{\tdh}}^{\tl\bar{\tl}  }  }   \}$.

Note that the CAS calculation can be systematically improved in the sense of configuration interactions (CI). Below, the above CAS(2,2) calculation will be extended to involve certain sets of configuration states. 

\subsubsection{CIS-1D}\label{sec:method:cis1d}
Recently, Teh et al have explored the notion of adding one carefully selected, double excitation state to a CIS wavefunction (or TD-DFT pseudo-wavefunction in the Tamm-Dancoff approximation). 
In the gas phase, this idea was proposed earlier by Maitra, Cave, and Burke long ago and was shown to improve excitation energies. In Ref. \cite{Teh2019}, Teh et al showed results for ethylene and stilbene, suggesting that this approach could recover both good excitation energies as well as an accurate $S0-S1$ crossing (with a correct topology for a conical intersection). The double excitation state was of the form $\ket{\Psi_{\tdh\bar{\tdh}}^{\tl\bar{\tl}} }$, where the active occupied and virtual orbitals, $\ket{\tdh}$ and $\ket{\tl}$, were chosen such that $\ev{H}{\Psi_{\tdh\bar{\tdh}}^{\tl\bar{\tl}}  }$ is minimized.

Here, we will adopt the idea of Ref. \cite{Teh2019} to study the AIM, which, unlike ethylene or stilbene, is not a closed system. Our ansatz for the optimized ground state will be obtained by minimizing $\ev{H}{\Psi_g}$ with 
\begin{align}\label{}
\ket{\Psi_g} = c_0 \ket{\Psi_0} +\frac{1}{\sqrt{2}} \sum_{jb} c_{j}^b \left( \ket{\Psi_j^b} + \ket{\Psi_{\bar{j}}^{\bar{b}}} \right) + \tilde{c} \ket{\Psi_{\tdh\bar{\tdh}}^{\tl\bar{\tl}}  }
\end{align}
where
\begin{align}\label{}
\ket{\tdh} &= \sum_{j} \ket{j} R_{j\tdh} \\
\ket{\tl} &= \sum_{b} \ket{b} Q_{b\tl}
\end{align}
are rotated active occupied and virtual orbitals yet to be determined.

Now, we would like to determine $\ket{\tdh}$ and $\ket{\tl}$ by an energetic minimization criterion (just as minimizing $\ev{H}{\Psi_{\tdh\bar{\tdh}}^{\tl\bar{\tl}}  }$ in Ref. \cite{Teh2019}). However, for our purposes, such minimization would not be effective for the same reason that choosing CAS active orbitals at a metal surface is impossible.
After all, for a system with a continuum of states near the Fermi level, the lowest-energy doubly-excited state is very likely to be a bath excitation, and this one double excitation will not yield a large correction. Thus, instead of minimizing $\ev{H}{\Psi_{\tdh\bar{\tdh}}^{\tl\bar{\tl}}  }$, we will simply adopt the active orbitals Eqs. \ref{eq:proj:h}-\ref{eq:proj:l}.

\subsubsection{CIS-ND}\label{sec:method:cisnd}
In the spirit of variational ansatz, one should expect that CIS-1D will be further improved if more configuration states are involved. Consider adding the following double excitation states to a CI Hamiltonian:
\begin{align}\label{}
\frac{1}{\sqrt{2}} \left(  \ket{\Psi_{\tdh\bar{\tj}}^{\tl\bar{\tb}}} + \ket{\Psi_{\bar{\tdh}\tj}^{\bar{\tl}\tb}} \right)
\end{align}
Here $\tj$($\tb$) is an arbitrary rotated occupied (virtual) orbital restricted only such that $(\tj\tb) \neq (\tdh\tl)$ (but $\tj$ or $\tb$ may equal to $\tdh$ or $\tl$ individually).
The total number of such double excitation states, plus the special double in CIS-1D, is $N_{occ} N_{vir} $. Let us denote the diagonalization of this CI Hamiltonian CIS-ND.

\subsubsection{MRCIS}
If we further allow  triple excitations of the form,
\begin{align}\label{}
\frac{1}{\sqrt{2}} \left(\ket{\Psi_{\tdh\bar{\tdh}\tj}^{\tl\bar{\tl}\tb}} +   \ket{\Psi_{\tdh\bar{\tdh}\bar{\tj}}^{\tl\bar{\tl}\bar{\tb}}} \right)
\end{align}
where $\tj\neq \tdh$ and $\tb \neq \tl$, one has essentially constructed a multi-reference configuration interaction singles (MRCIS) calculation with three reference configurations $\ket{ \Psi_0 }$ , $\ket{ \Psi_{\tdh}^{\tl}  }_s$and $\ket{ \Psi_{\tdh \bar{\tdh}}^{\tl\bar{\tl}  }  }$.  For each of these configurations, we allow single excitation on top, so that the size of the total basis set is about $3N_{occ} N_{vir}$ (the actual size is slightly less due to double counting).

\subsection{Unrestricted mean-field}
For the AIM, a non-magnetic ($\ev{n_{\uparrow}} = \ev{n_{\downarrow}}$), single-determinant  ($\ev{n_{\uparrow} n_{\downarrow}} = \ev{n_{\uparrow}} \ev{n_{\downarrow}}$) state is energetically disfavored in the presence of a large $U$ parameter. Consequently, a variational closed-shell, single-determinant ground state can be qualitatively wrong in the strongly-correlated regime, and thus we should not expect standard mean-field theory (Eqs. \ref{eq:gs:mf}-\ref{eq:pop}) to be very accurate.
Now, for all of the CI methods above, we go beyond the mean-field limitation, and yet the optimization of the ground state is still performed {within a subspace of non-magnetic configuration states}.
Of course, one can imagine a cheaper alternative: one can seek a lower energy, variational ground state not by relaxing the mean-field property, but instead by relaxing the non-magnetic constraint.
This ansatz essentially leads to an unrestricted mean-field calculation.
Consider an unrestricted Slater determinant,
\begin{align}\label{}
\ket{\Psi} = \ket{i_{1}\bar{i}_{2}j_{1} \bar{j}_{2}\ldots   }
\end{align}
Following the variational principle, spin-up and spin-down orbitals are now determined by the following two coupled self-consistent equations:
\begin{align}\label{}
(h+U \bar{n}_0 P) \ket{i} &= \lambda_i \ket{i} \label{eq:umf:up}\\
(h+U n_0 P) \ket{\bar{j}} &= \bar{\lambda}_j \ket{\bar{j}}
\end{align}
where 
\begin{align}\label{}
n_0 &\equiv \sum_{i \in\Psi}\abs{\ip{i}{d}}^2 \\
\bar{n}_0 &\equiv \sum_{\bar{j} \in\Psi}\abs{\ip{\bar{j}}{d}}^2 \label{eq:umf:ndown}
\end{align}
By solving Eqs. \ref{eq:umf:up}-\ref{eq:umf:ndown}, one obtains an unrestricted mean-field solution to the AIM. Below, we will use CIS-1D, CIS-ND, MRCIS, CAS(2,2), and unrestricted mean-field to solve for the ground state of the AIM, and, by comparing with NRG, assess how well these methods perform in an open Hamiltonian (rather than the more standard case of a small molecule).

\subsection{Diabatization}\label{sec:method:diab}
As mentioned above, to model nonadiabatic dynamics, a diabatic representation will be helpful to establish a chemical picture and to serve as a basis for model calculations (sometimes computational, sometimes analytical). 
As such, we would very much like a reduced diabatic picture of the electronic structure within the AIM.
Unfortunately, however, our current situation differs from the conventional diabatization problem in molecular systems in that, while 
we may expect three diabatic PESs representing zero-, one- and two-electron occupation, these states must be extracted from a continuum of adiabatic states; the standard notion of adiabatic-to-diabatic transformations\cite{Baer1975, Baer1980,Pacher1988,Pacher1991,Ruedenberg1993,Ruedenberg1997,Truhlar2001,Truhlar2002,Truhlar2003,Subotnik2008,Subotnik2009} are not easily applied.
For this reason, we will now employ a different, projection-based strategy to perform such a diabatization.

To better understand our approach, let us begin by rewriting the identity\cite{Schonhammer1975}:
\begin{align}\label{eq:breakI}
1 =  (1-\nup)(1-\ndown) +  \left( \nup(1-\ndown) + \ndown(1-\nup) \right)  + \nup\ndown \equiv P_0 + P_1 + P_2
\end{align}

For any many-body state $\ket{\Psi}$, $\{ P_0 \ket{\Psi}, P_1 \ket{\Psi}, P_2 \ket{\Psi} \}$ are the three components of $\ket{\Psi}$ with zero-, one- and two-electron population (up to some normalization constants), and in principle this would be an excellent set of diabtic states. 
However, since neither the CIS-1D basis, nor the CIS-ND basis, nor the MRCIS basis is an invariant subspace of $P_M (M=0,1,2)$\cite{footnote:diab}, 
projection by $P_M$ on a CIS-1D/CIS-ND/MRCIS wavefunction
will bring one given vector outside the space of optimization, and therefore evaluating matrix elements of $P_M H P_{M'}$ would be laborious. 
Thus, a slightly simple approach would be preferable.

To that end, let us define
\begin{align}\label{}
\tilde{P}_M \equiv Q P_M Q
\end{align}
where $Q$ is the projection operator onto the optimization basis. For instance,
\begin{align}\label{}
Q^{(CIS-1D)} \equiv \dyad{\Psi_0}{\Psi_0} + \frac{1}{2}\sum_{jb} \left( \ket{\Psi_j^b} + \ket{\Psi_{\bar{j}}^{\bar{b}}} \right) \left( \bra{\Psi_j^b} + \bra{\Psi_{\bar{j}}^{\bar{b}}} \right) + \dyad{\Psi_{\tdh\bar{\tdh}}^{\tl\bar{\tl}}}{\Psi_{\tdh\bar{\tdh}}^{\tl\bar{\tl}}} 
\end{align}
If we denote the ground state $\ket{\Psi_g}$, then we can also define the following diabatic state
\begin{align}\label{}
\ket{\tilde{\Xi}_M} \equiv  \frac{\tilde{P}_M \ket{\Psi_g}}{\sqrt{\ev{\tilde{P}_M^2}{\Psi_g}}} 
\end{align}
where $\ket{\tilde{\Xi}_M}$ is a potentially diabatic state whose system population is approximately $M$. 

While the set $\{\ket{\tilde{\Xi}_M}\}$ would appear to be a natural basis of  diabatic states, one important drawback is that this set is not orthonormal.
To yield an orthonormal basis while preserving charge character as much as possible, a L\"owdin orthogonalization\cite{Lowdin1950, Lowdin1970, Carlson1957, Fischer1965}
can be applied,
\begin{align}\label{eq:method:rawdiab}
\ket{\Xi_M} \equiv \sum_{M'} \ket{\tilde{\Xi}_{M'}} (S^{-1/2})_{M'M}
\end{align}
where $S$ is the overlap matrix, i.e. $S_{MM'} = \ip{\tilde{\Xi}_M}{\tilde{\Xi}_{M'}}$. The full diabatic Hamiltonian $\mathcal{H}$ is then readily obtained by
\begin{align}\label{}
\mathcal{H}_{MM'} = \mel{\Xi_M}{H}{\Xi_{M'}}
\end{align}
Below, we will report on the relevant set of diabatic states $\{ \ket{\Xi_M}\}$ as found for the AIM.

\section{Results} \label{sec:result}
We will now present results for the various methods in Table \ref{table:CI} as far as reproducing the correct physics behind the AIM.

\begin{figure}
	\begin{subfigure}[t]{.95\textwidth}
		\includegraphics{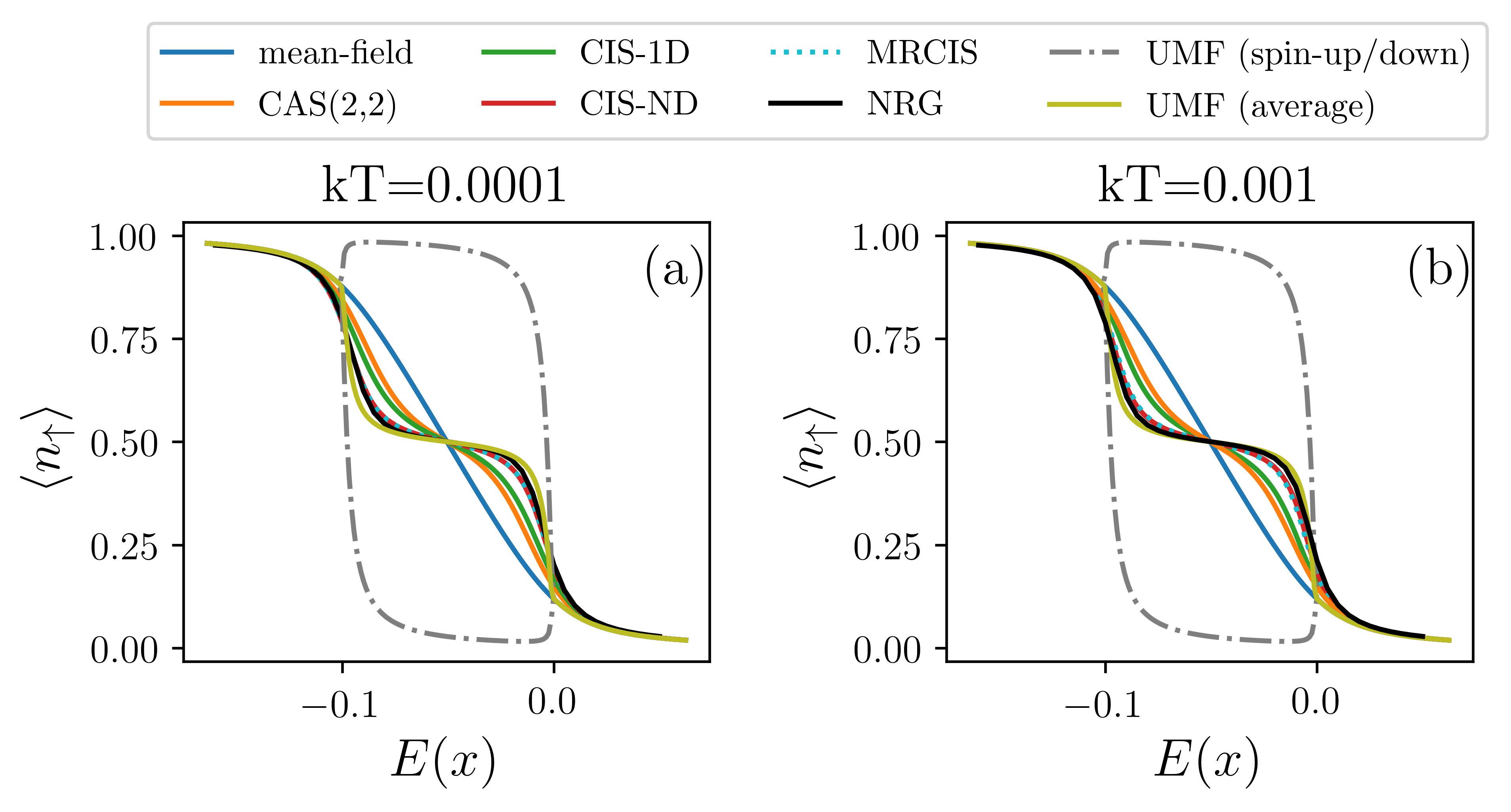}
	\end{subfigure}	

	\begin{subfigure}{.5\textwidth}
		\includegraphics{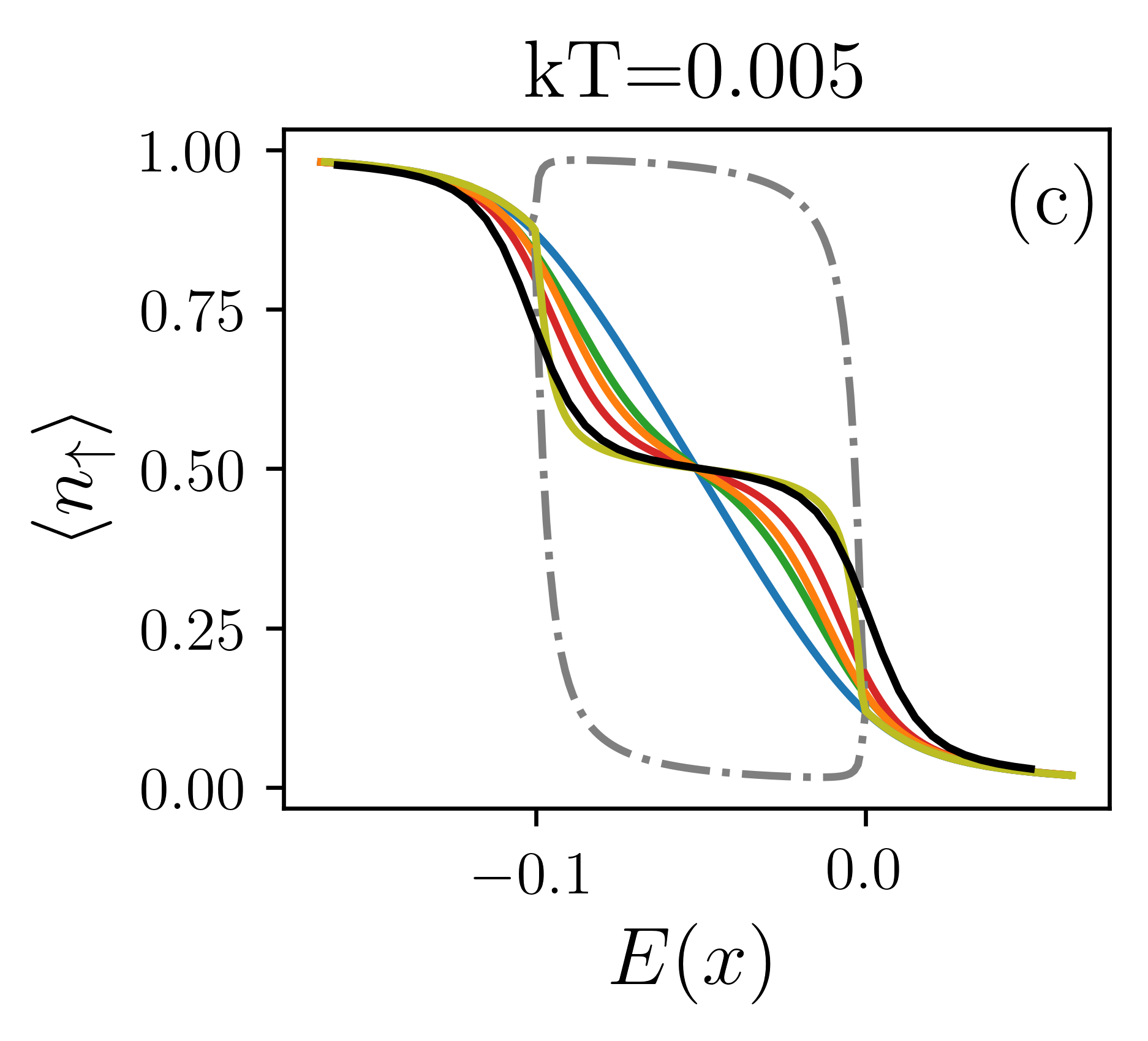}
	\end{subfigure}	
	\begin{minipage}{.45\textwidth}
		\caption{System population as a function of system on-site energy $E(x)$ (see Eq. \ref{eq:H:AH}). The bath state energies range from -0.2 to 0.2 with spacing 0.001 (a and b) and 0.003 (c). The temperatures are (a) 0.0001, (b) 0.001 and (c) 0.005. Other parameters are $U=0.1$, $\Gamma=0.01$, $\mu=0$. NRG is used as the benchmark. The restricted mean-field (MF) ground state entirely fails to capture the electronic correlation, while the unrestricted calculation (UMF) over-corrects the impurity population and introduces artificial continuities between the correlated and non-correlated regime. A three-state CAS(2,2) optimization can introduce significant correction to the mean-field population. CIS-ND and MRCIS reproduce NRG results quite well when $ kT \lesssim \Delta E \ll \Gamma$. All the CIS-based methods deteriorate when $\Delta E < kT \lesssim \Gamma$.}
		\label{fig:n_imp}
	\end{minipage}
\end{figure}

Fig. \ref{fig:n_imp} plots the impurity population vs. on-site energy for the various methods at different temperatures. A NRG calculation is used as the exact benchmark.
In Fig. \ref{fig:n_imp}a  we set $U = 10\Gamma = 10^2 \Delta E = 10^3 kT$, where $\Delta E$ is the bath spacing. The CIS-1D optimized ground state exhibits a moderate correction to the restricted mean-field ground state, but is still far from the NRG result.
Interestingly, a CAS(2,2) optimization accounts for a large portion of the CIS-1D correction.
The CIS-ND and MRCIS optimized ground states clearly display population plateaus almost as does the exact NRG, indicating that these methods include a significant correction to the restricted mean-field ground state.
As far as the unrestricted mean-field (UMF) ground state, the UMF population coincides with the restricted mean-field population in the two non-correlated regions ($E(x)<\mu-U$ and $E(x)>\mu$), and displays a clear population plateau in between (actually even broader than NRG's). However, discontinuities  exist at the intersection of the correlated and non-correlated regions, similar to the Coulson-Fischer point\cite{Coulson1949} in standard unrestricted Hartree-Fock calculations. Such discontinuities are  artifacts that result from the fact that the method is limited to a single Slater determinant, and this limitation can lead to unphysical artifacts or divergences of quantities related to dynamics, e.g., the electronic friction\cite{Trail2003} (see the discussion section).

In Figs. \ref{fig:n_imp}b and \ref{fig:n_imp}c, the temperature is raised so that $kT/\Gamma = 0.1$ and $0.5$ respectively.
While the performance of CI-based methods is apparently unaffected in Fig. \ref{fig:n_imp}b, the performance significantly deteriorates in Fig. \ref{fig:n_imp}c, where a significant portion of the excitation spectrum is involved in the Boltzmann average ($kT/\Gamma=0.5$).
This failure should not be surprising, as the configuration states included in the diagonalization  by no means reflect the underlying density of states, and therefore the excited state DoS for the CI-based methods above are not reliable.
As a consequence, the high-$T$ Boltzmann-weighted average is also biased, leading to the deterioration of the CI-based methods in Fig. \ref{fig:n_imp}c at large temperature.

\begin{figure}
	\begin{subfigure}[t]{.95\textwidth}
		\includegraphics{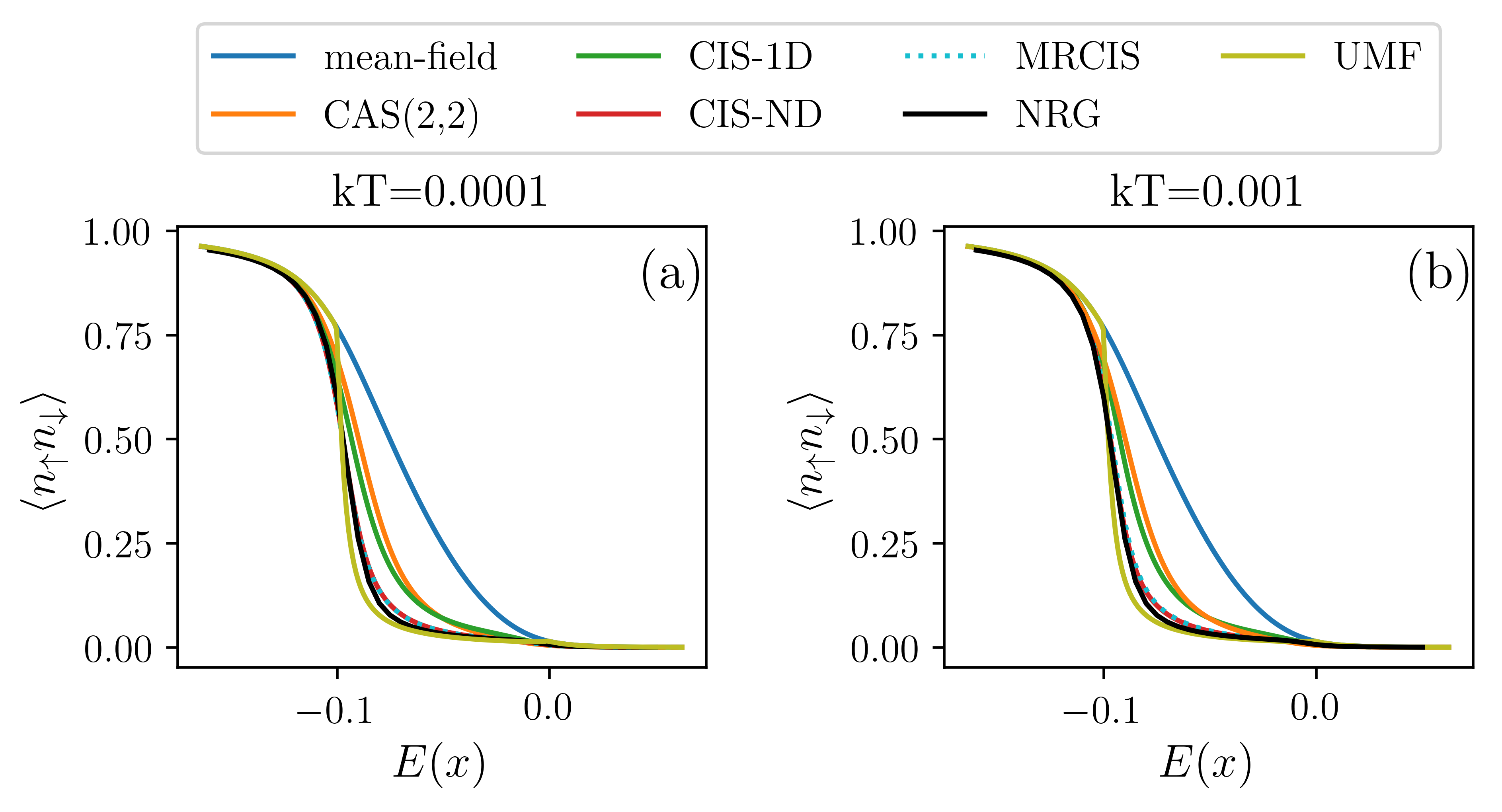}
	\end{subfigure}	
	
	\begin{subfigure}{.55\textwidth}
		\includegraphics{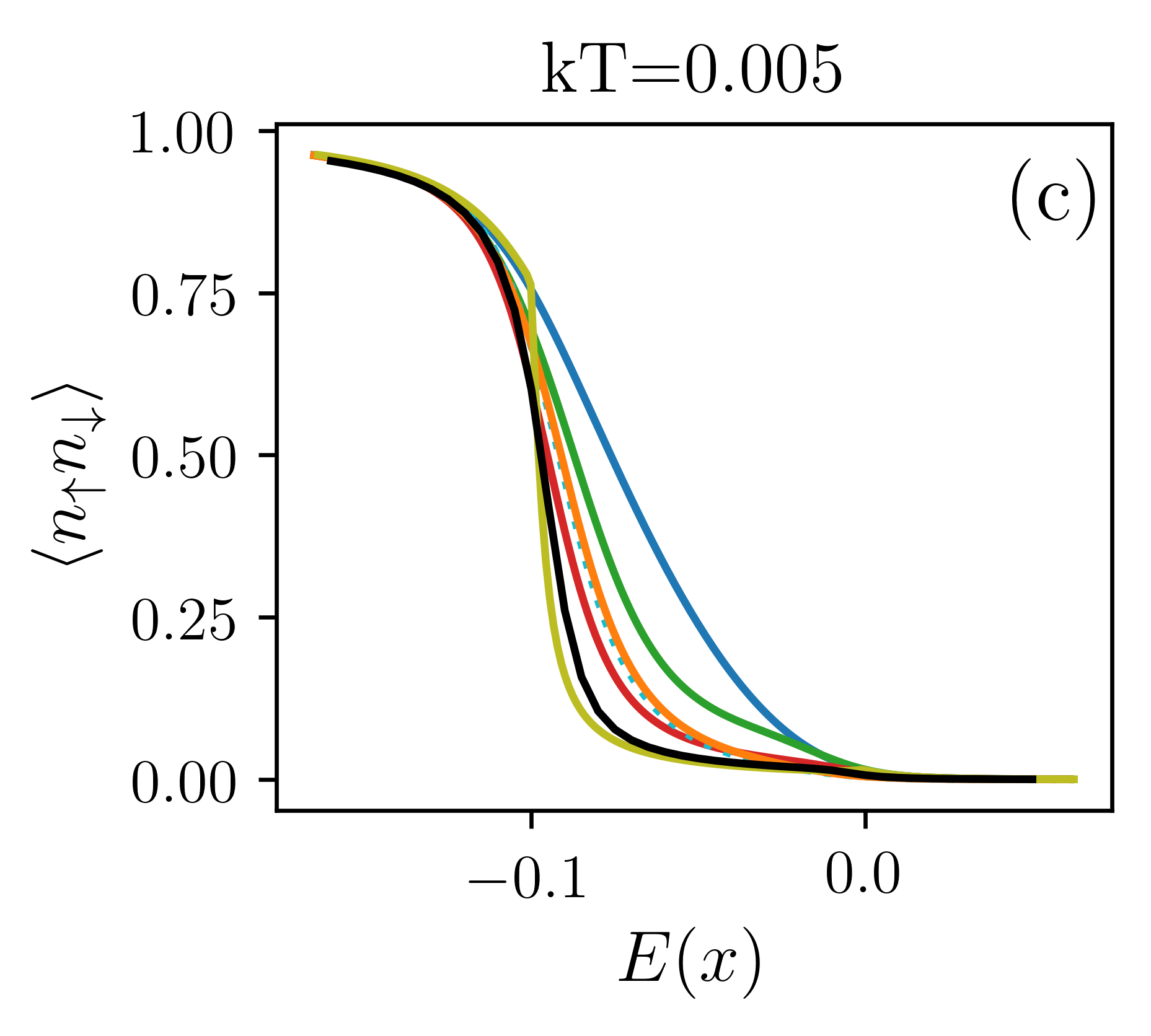}
	\end{subfigure}	
	\begin{minipage}{.4\textwidth}
		\caption{On-site repulsion as a function of system on-site energy $E(x)$. The parameters are the same as those in Fig. \ref{fig:n_imp}. The repulsion is nearly zero between $-0.1$ and 0, where the impurity population is around 1, demonstrating the multi-reference nature in this region. In all cases CI methods can perform well for small enough temperatures.}
		\label{fig:nn_imp}
	\end{minipage}
\end{figure}

Next, in Fig. \ref{fig:nn_imp}, the on-site repulsion is plotted as a function of on-site energy with the same parameter set as used in Fig. \ref{fig:n_imp}.
In all cases, the NRG repulsion rapidly drops to about zero at the point where $E(x) = \mu-U$, which verifies the multi-reference character as described by Eq. \ref{eq:tworef}.
The relative performance of the CI-based methods agrees with their impurity populations as in Fig. \ref{fig:n_imp}; for small temperature $T$, CI-based methods can ideally recover $\ev{\nup \ndown}$ for a large enough CI basis.

\begin{figure}
	\includegraphics[width=6.3in]{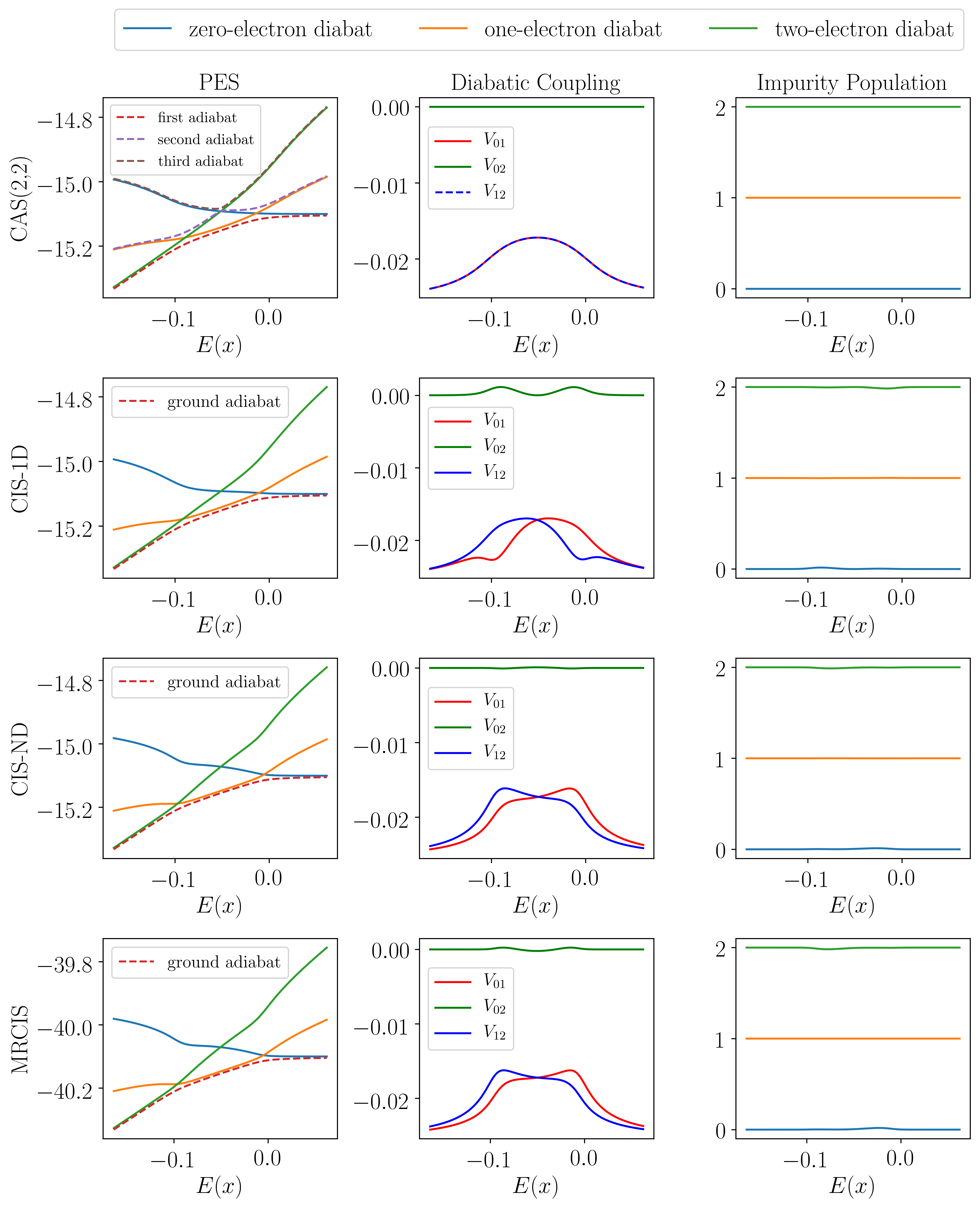}
	\caption{Diabatic PESs, couplings and impurity populations from CIS-1D, CIS-ND, MRCIS and CAS(2,2). The zero-, one- and two-electron diabats are plotted in blue, orange and green respectively. The projection diabatization for the CAS(2,2) algorithm yields exact integer-occupation quasi-diabatic states, since the basis is an invariant subspace of $P_M$. For the other three methods, the system populations of diabatic states are not exact integer, yet the deviations are very small. The diabatic couplings of MRCIS do not significantly differ from those of CIS-ND. }
	\label{fig:diab}
\end{figure}

Finally, in Fig. \ref{fig:diab}, we plot the AIM quasi-diabatic PESs and diabatic couplings as defined in Sec. \ref{sec:method:diab}.
Encouragingly, the projection-based diabatization yields similar diabatic Hamiltonians and impurity populations in all of the  different levels of electronic structure. 
In the three-state (CAS(2,2)) calculation, the impurity population is an exact integer because the three-state basis is an invariant subspace of $P_M$ ($\tilde{P}_M \equiv Q P_M Q = P_M$); for the other CI calculations, the system populations are close to (but not exactly) integer values. 
Moreover, for the different levels of CI theory, the exact positions of the diabatic crossings differ only slightly.
In general, the CIS-ND diabatized Hamiltonian does not differ significantly from that of MRCIS.

Overall, the takeaway message from Figs. \ref{fig:n_imp}-\ref{fig:diab} is clear: if the temperature is small enough ({but not so low as to enter the Kondo regime}), one can clearly use carefully-designed CI-based approaches to model the electronic structure of the AIM.


\section{Discussion}\label{sec:discussion}
\begin{figure}
	\begin{subfigure}{.55\textwidth}
		\includegraphics[width=3.2in]{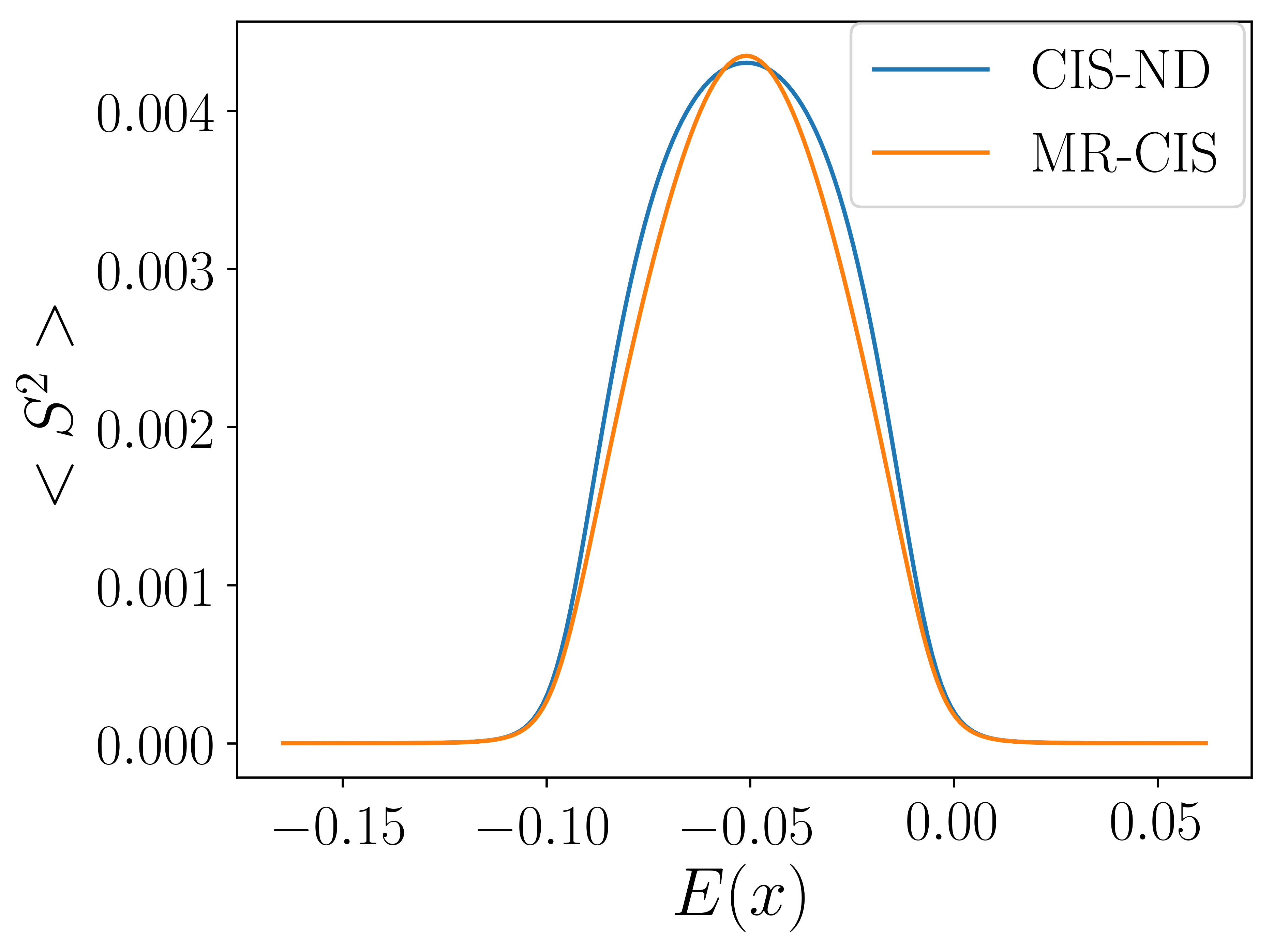}
	\end{subfigure}
	\begin{minipage}{.4\textwidth}
		\caption{ {$\ev{S^2}$ vs. on-site energy for CIS-ND and MR-CIS ground state. The configuration states introduced by CIS-ND are not spin-adaptive, thus spin contamination occurs in the correlated regime. Nevertheless, the contamination is insignificant (less than 0.005). } } 
		\label{fig:S2}
	\end{minipage}
\end{figure}

{
\subsection{Spin Contamination}
One drawback of the CIS-ND and MR-CIS approaches above is that they are not spin-adaptive; the double excitations introduced by CIS-ND are not pure singlets. However, since the double excitations are introduced mostly to relax the orbitals from the excited single-excitation singlet configurations, these states themselves do not lead to a large spin contamination.
Fig. \ref{fig:S2} plots the expectation value $\ev{S^2}$ for the CIS-ND and MR-CIS optimized ground state. The spin contamination is less than 0.005 throughout the correlated regime and effectively zero elsewhere.
}

\subsection{Electronic Friction}
In studying the electronic structure of the AIM, one of the most important question is: how we will add in possibly nonadiabatic dynamics when nuclei start moving?
Of course, in general, modeling nonadiabatic dynamics on metal surfaces is a non-trivial task. 
However, in the limit of fast electronic equilibration, it is known that nonadiabatic dynamics can be reduced to Langevin dynamics\cite{Tully1995}, where the nonadiabatic effects are incorporated into the electronic friction. 

\begin{figure}
	\includegraphics[width=6.3in]{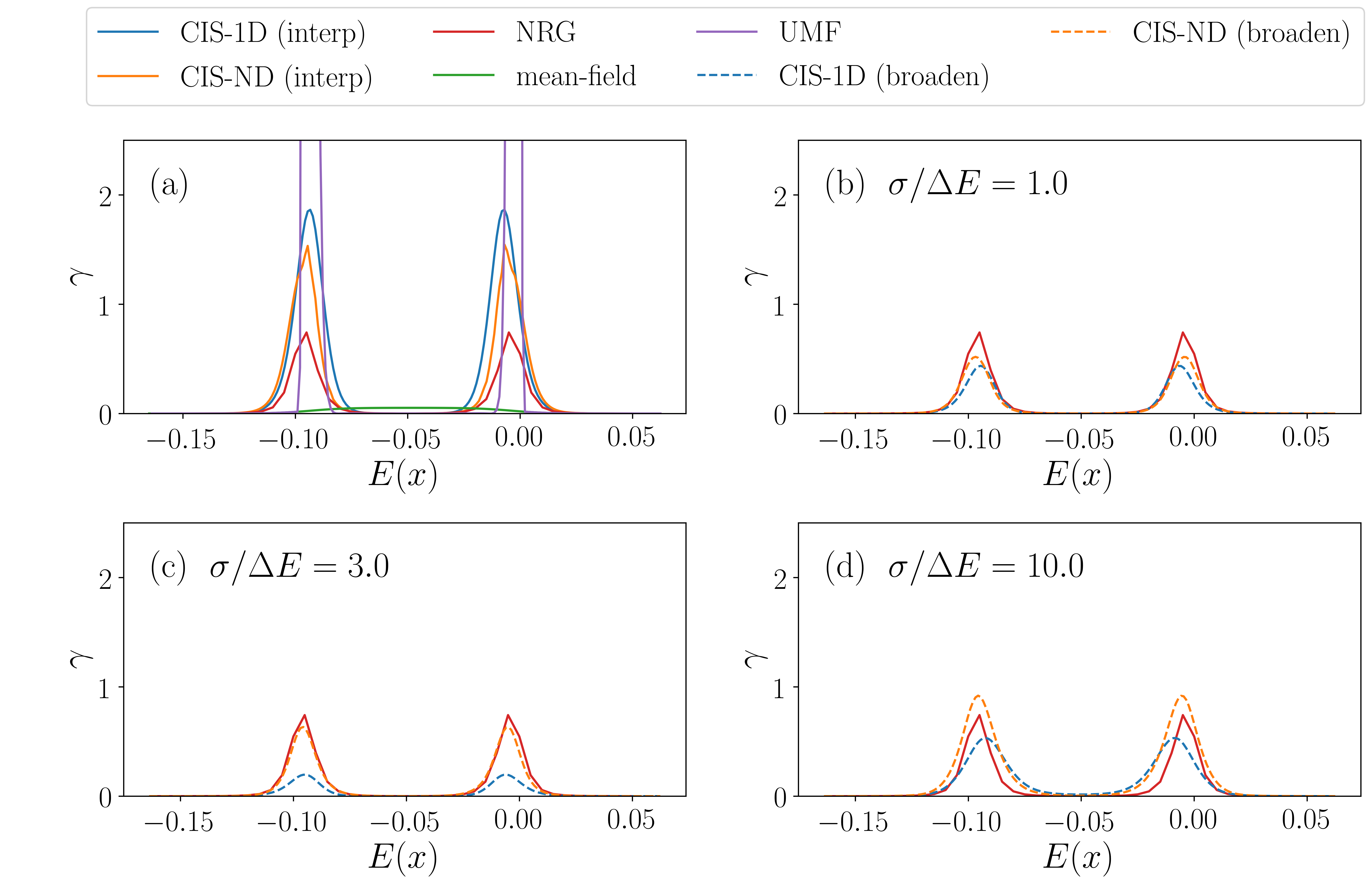}
	\caption{ Electronic friction ($\gamma$) vs impurity on-site energy $(E(x))$ according to CIS-1D, CIS-ND, mean-field theory and NRG. The parameters are the same as those in Fig. \ref{fig:n_imp}b. Unlike restricted mean-field theory, CIS-1D and CIS-ND can qualitatively capture the double-peaked pattern; yet their quantitative value is not reliable here, due to both numerical instability as well as systematic errors in the DoS. The CIS-based friction in the upper left panel is computed by an interpolation method, while in the remaining panels friction tensors are computed by broadening the delta functions ($\sigma$ is the width of gaussian and $\Delta E$ is the bath spacing). Note that the friction tensors computed by interpolation are not accurate while the friction tensors computed by broadening do not converge with $\sigma$; see Appendix A for details. The unrestricted mean-field friction diverges at the Coulson-Fischer point. Note that we cannot report CAS(2,2) friction tensor here because friction requires a continuum of states, and for the CAS(2,2) calculation we do not have such a set of bath states.
	}
	\label{fig:fric}
\end{figure}

Now, Ref. \cite{Dou2017prl} has demonstrated that, in the case of moderate to high temperature and $U\gg \Gamma$ limit, the exact electronic friction in the AIM peaks at two positions, exactly those positions where electron transfer should occur.
And, as also shown by Ref. \cite{Dou2017prl}, these two peaks are not captured by the restricted mean-field theory\cite{Dou2017prl}. 
With this background in mind, in Fig. \ref{fig:fric}, we report CIS-1D and CIS-ND friction coefficients as compared against NRG and mean-field friction.
Encouragingly, both CIS-1D and CIS-ND qualitatively recover the double-peaked pattern at exactly the position in space where the proposed diabats cross.
Unfortunately, however, their quantitative values in the figure are not reliable. On the one hand, the electronic friction depends on the density of states, and none of the CIS-based methods can capture the true DoS; on the other hand,
the Dirac delta functions in the friction formula\cite{Dou2017prl}
\begin{align}\label{eq:ef}
\gamma_{\mu\nu} = \frac{ \pi\hbar\beta }{2} \sum_{ IJ } \mel{I}{ \delta \hat{F}_{\mu} }{J} \mel{J}{\delta \hat{F}_{\nu}}{I}  \frac{e^{-\beta E_I} + e^{-\beta E_J} }{Z} \delta(E_I-E_J)
\end{align}
induces numerical instabilities: if one simply replace the delta functions with some broadening functions (like gaussian), the number of bath states should be sufficiently dense compared with $kT$ and $\Gamma$ in order to converge the friction (which would be hard).
While an interpolation scheme has been proposed to alleviate this issue in single-electron orbital framework \cite{Jin2019}, this method does not seem to reliably extend to a many-body framework. For computational details, see Appendix \ref{apdx:fric}.
Lastly, we also note that, in Fig. \ref{fig:fric}, the unrestricted mean-field friction does produce two peaks, but the friction diverges at the Coulson-Fischer points because of the abrupt changes in the slopes of impurity population. Similar artifacts in the friction tensor have been observed earlier by Trail et al\cite{Trail2003}.

Overall, the details in Fig. \ref{fig:fric} would suggest that, despite the success of CI-based methods at recovering the correct qualitative electronic friction of the AIM, modeling nonadiabatic molecular dynamics with electronic friction does not appear simple or straightforward, and finding other methods will be helpful, likely based on surface hopping\cite{Tully1990}.
In any event, it is known that electronic friction can fail (even with the correct friction tensor) and our past experience suggests that surface hopping is a better starting point for exploring complicated dynamics at surfaces\cite{Miao2017}.

\section{Conclusion}\label{sec:conclusion}
In the present work, we have studied the Anderson impurity model with configuration interaction electronic structure methods.
While a restricted mean-field calculation fails to capture the effect of electron-electron effect, we find that, with one special double excitation added to the CIS calculation, we are able to partially capture static correlation and partially recover the correct ground state impurity population.
With a total of about $2N_{occ} N_{vir}$ configuration states  (CIS-ND), the ground state impurity population can recover almost exactly the NRG results. However, such CI methods cannot give the correct temperature dependence, as these truncated CI methods cannot generate the right density of states.

Looking forward to dynamics applications, we have proposed a projection-based diabatization scheme to obtain a diabatic representation of the total system.
Motivated by chemical intuition, this diabatization makes use of the impurity on-site charge as the criterion for separating components of different chemical character in the ground state.
As such, this diabatization differs from traditional approaches based on generating an adiabatic-to-diabatic transformation.
Furthermore, unlike diabatization in the gas phase\cite{Baer1975, Baer1980,Pacher1988,Pacher1991,Ruedenberg1993,Ruedenberg1997,Truhlar2001,Truhlar2002,Truhlar2003,Subotnik2008,Subotnik2009}, here we generate diabatic states only from the ground state (as opposed to using a collection of adiabatic states).
Admittedly, generating diabatic states from the ground state alone might appears dubious, even at low temperatures, and yet our belief is that this approach should be suitable for capturing charge-transfer dynamics of molecules on surfaces, e.g. the famous NO on gold surface experiments\cite{Wodtke2000}. Further investigation will be necessary to evaluate the overall effectiveness and accuracy of this diabatic representation.

Now, with regard to applicability, we should emphasize that all the calculations in the present paper have been limited to the AIM.
Nevertheless, the approaches introduced in Sec. \ref{sec:method} should be all applicable to \textit{ab initio} calculations as post-DFT methods, where we will interpret the Kohn Sham wavefunction as a real wavefunction (just as one did in Ref. \cite{Teh2019}).
And yet one key step remains before we can apply the present approach to realistic systems of molecules on metal surfaces: 
molecules will have more than one orbital, as opposed to the single-impurity model used in the paper.
And thus, we will need a predicative strategy (ideally a black-box algorithm) for finding the active $\tilde{h}$ and $\tilde{l}$ orbitals for an \textit{ab initio} calculation.

Finally, simulating molecular nonadiabatic dynamics on metal surfaces in the presence of electron-electron correlation remains a challenging task. 
With a system-wide few-level diabatic representation (e.g., as proposed in the present work), one should in principle be able to use semiclassical methods (e.g. the fewest-switches surface hopping algorithm\cite{Tully1990}) to simulate the dynamics.
And yet the validity of Tully's algorithm in this scenario is unclear, since this method would not reflect the correct electronic equilibration.
For this reason, Shenvi et al\cite{Tully2009I} originally developed the independent electron surface hopping model.
Nevertheless, for all of its successes\cite{Tully2009M}, the IESH model relies on an independent electrons and cannot directly include electron-electron repulsion (as would be found in a realistic \textit{ab initio} Hamiltonian); finite temperature effects are also difficult\cite{Miao2019}.

Alternatively, assuming an adsorbate-substrate separation, a broadened classical master equation\cite{Dou2016BCME,Dou2017BCME} would appear to be a good candidate for modeling the molecule-metal problem in the molecular diabatic picture. However in practice, obtaining the correct hybridization $\Gamma$ in the case of a many-body electronic structure framework is not trivial. In the future, finding the appropriate pairing between correlated electronic structure calculations and semiclassical nonadiabatic dynamical methods is an exciting goal.

\section{Acknowledgment}
This work was supported by the U.S. Air Force Office of Scientific Research (USAFOSR) AFOSR Grants No. FA9550-18-1-0497 and FA9550-18-1-0420. Computational support was provided by the High Performance Computing Modernization Program of the Department of Defense.

\appendix
\section{Computational Details of Electronic Friction}\label{apdx:fric}
In this appendix, we will introduce two ways of evaluating Eq. \ref{eq:ef} numerically.

\subsection{Gaussian Broadening}
A straightforward and common treatment of the Dirac delta functions in Eq. \ref{eq:ef} is to replace them with some broadening functions of finite width, e.g.
\begin{align}\label{}
\delta(E_I- E_J) \rightarrow \frac{1}{\sigma\sqrt{2\pi}} e^{-\frac{(E_I-E_J)^2}{2\sigma^2}}
\end{align}
where $\sigma$ is a parameter that controls the broadening width. Ideally, one would expect that $\gamma_{\mu\nu}(\sigma)$ can plateau at a wide range of $\sigma$ value, which is the converged result. Nevertheless, such convergence is guaranteed only when $\Delta E \ll \Gamma$ and $\Delta E \ll kT$ are both satisfied ($\Delta E$ is the bath spacing near the Fermi level). Without a sufficiently dense bath, a broadening method is very likely to fail.
Note that the interpolated friction tensors in Fig. \ref{fig:fric} do not converge with broadening factor, $\sigma$.

\subsection{Interpolation}
The friction can also be evaluated with an interpolation scheme similar to the one used in Ref. \cite{Jin2019}.
First, note that Eq. \ref{eq:ef} can be recast into an integral form
\begin{align}\label{eq:ef:int}
\gamma_{\mu\nu} = -\pi \hbar \int \dd\epsilon \text{Tr}\left( \delta \hat{F}_{\mu} P(\epsilon) \delta \hat{F}_\nu P(\epsilon) \right) \pdv{p(\epsilon)}{\epsilon} 
\end{align}
where $P(\epsilon) \equiv \delta(\epsilon - H)$ and $p(\epsilon) \equiv \exp(-\beta \epsilon)/Z$. Now, assuming the eigenspectrum of $H$ is \textit{non-degenerate} at every eigen-energy $E_I$, we can deduce that
\begin{align}\label{}
 \text{Tr}\left( \delta \hat{F}_{\mu} P(E_I) \delta \hat{F}_\nu P(E_I) \right) = \text{Tr}\left( \delta \hat{F}_{\mu} P(E_I) \right) \text{Tr} \left( \delta \hat{F}_\nu P(E_I) \right)
\end{align}
Let us define the cumulative sum function
\begin{align}\label{eq:cumsum}
I_\mu (\epsilon) \equiv \sum_{J}^{E_J < \epsilon} \ev{\delta F_\mu}{J}
\end{align}
Obviously,
\begin{align}\label{}
\dv{I_\mu}{\epsilon} = \text{Tr}\left( \delta \hat{F}_{\mu} P(\epsilon) \right)
\end{align}
Therefore, Eq. \ref{eq:ef:int} can be written as
\begin{align}\label{eq:ef:interp}
\gamma_{\mu\nu} = -\pi\hbar \int \dd \epsilon \dv{ I_{\mu}}{\epsilon} \dv{ I_{\nu}}{\epsilon} \pdv{p(\epsilon)}{\epsilon}
\end{align}
An interpolation method can be established as follows, assuming that one has already obtained the eigenpairs $\{\ket{I}, E_I \}$:
\begin{enumerate}
	\item[i.] Find $\ev{\delta F_\mu}{I} \equiv -\ev{\partial_{\mu}H}{I} + \ev{\partial_\mu H}$. In the specific model Hamiltonian (Eq. \ref{eq:H:AH}), this can be simplified to $-\left(\sum_{\sigma}\ev{ n_\sigma}{I} - \ev{n_\sigma}\right)$. In a realistic calculation, this quantity can be computed by density functional perturbation theory.
	\item[ii.] Compute the cumulative sum function $I_\mu(E_I)$ according to Eq. \ref{eq:cumsum};
	\item[iii.] Fit $I_{\mu}(E_I)$ to some smooth function $\tilde{I}_{\mu}(\epsilon)$. While in the single-electron picture one may pick a functional form based on the analytical expression in the wide-band limit, a many-electron counterpart is not available. In practice, the fallback method is the cubic spline over a smoothened data set.
	\item[iv.] Numerically compute Eq. \ref{eq:ef:interp}.
\end{enumerate}
There are, however, two factors that undermine the quality of this interpolation scheme. First, the assumption of non-degeneracy is more problematic in the many-body case than in the one-body case. Second, the curve fitting is prone to numerical instabilities: unlike the situation in single-electron picture whereby the energy range of interest (a few $kT$ around the Fermi level) usually lies within the total energy range, the current thermally dominant range actually lies near the lower bound of the total spectrum, essentially making interpolation an extrapolation. Due to the two reasons above, the interpolation method does not so far appear very reliable.


\begin{thebibliography}{10}
	
	\bibitem{Rodriguez1992}
	Jos{\'e}~A. Rodriguez and D.~Wayne Goodman.
	\newblock The nature of the metal-metal bond in bimetallic surfaces.
	\newblock {\em Science}, 257(5072):897--903, 1992.
	
	\bibitem{Bunermann2015}
	Oliver B{\"u}nermann, Hongyan Jiang, Yvonne Dorenkamp, Alexander Kandratsenka,
	Svenja~M. Janke, Daniel~J. Auerbach, and Alec~M. Wodtke.
	\newblock Electron-hole pair excitation determines the mechanism of hydrogen
	atom adsorption.
	\newblock {\em Science}, 350(6266):1346--1349, 2015.
	
	\bibitem{Nienhaus1999}
	H.~Nienhaus, H.~S. Bergh, B.~Gergen, A.~Majumdar, W.~H. Weinberg, and E.~W.
	McFarland.
	\newblock Electron-hole pair creation at ag and cu surfaces by adsorption of
	atomic hydrogen and deuterium.
	\newblock {\em Phys. Rev. Lett.}, 82:446--449, Jan 1999.
	
	\bibitem{Gergen2001}
	Brian Gergen, Hermann Nienhaus, W.~Henry Weinberg, and Eric~W. McFarland.
	\newblock Chemically induced electronic excitations at metal surfaces.
	\newblock {\em Science}, 294(5551):2521--2523, 2001.
	
	\bibitem{Nitzan2003}
	Abraham Nitzan and Mark~A. Ratner.
	\newblock Electron transport in molecular wire junctions.
	\newblock {\em Science}, 300(5624):1384--1389, 2003.
	
	\bibitem{Ratner2013}
	Mark Ratner.
	\newblock A brief history of molecular electronics.
	\newblock {\em Nature nanotechnology.}, 8(6):378,381, 2013-6.
	
	\bibitem{Gautier2015}
	Sarah Gautier, Stephan~N. Steinmann, Carine Michel, Paul Fleurat-Lessard, and
	Philippe Sautet.
	\newblock Molecular adsorption at pt(111). how accurate are dft functionals?
	\newblock {\em Phys. Chem. Chem. Phys.}, 17:28921--28930, 2015.
	
	\bibitem{Wellendorff2015}
	Jess Wellendorff, Trent~L. Silbaugh, Delfina Garcia-Pintos, Jens~K. N{\o}rskov,
	Thomas Bligaard, Felix Studt, and Charles~T. Campbell.
	\newblock A benchmark database for adsorption bond energies to transition metal
	surfaces and comparison to selected dft functionals.
	\newblock {\em Surface Science}, 640:36 -- 44, 2015.
	\newblock Reactivity Concepts at Surfaces: Coupling Theory with Experiment.
	
	\bibitem{Grafenstein2000}
	J{\"u}rgen Gr{\"a}fenstein and Dieter Cremer.
	\newblock Can density functional theory describe multi-reference systems?
	investigation of carbenes and organic biradicals.
	\newblock {\em Phys. Chem. Chem. Phys.}, 2:2091--2103, 2000.
	
	\bibitem{Hedin1965}
	Lars Hedin.
	\newblock New method for calculating the one-particle green's function with
	application to the electron-gas problem.
	\newblock {\em Phys. Rev.}, 139:A796--A823, Aug 1965.
	
	\bibitem{Newns1969}
	D.~M. NEWNS.
	\newblock Self-consistent model of hydrogen chemisorption.
	\newblock {\em Phys. Rev.}, 178:1123--1135, Feb 1969.
	
	\bibitem{Gull2011}
	Emanuel Gull, Andrew~J. Millis, Alexander~I. Lichtenstein, Alexey~N. Rubtsov,
	Matthias Troyer, and Philipp Werner.
	\newblock Continuous-time monte carlo methods for quantum impurity models.
	\newblock {\em Rev. Mod. Phys.}, 83:349--404, May 2011.
	
	\bibitem{Bulla1997}
	R~Bulla, Th~Pruschke, and A~C Hewson.
	\newblock Anderson impurity in pseudo-gap fermi systems.
	\newblock {\em Journal of Physics: Condensed Matter}, 9(47):10463--10474, nov
	1997.
	
	\bibitem{Bulla2008}
	Ralf Bulla, Theo~A. Costi, and Thomas Pruschke.
	\newblock Numerical renormalization group method for quantum impurity systems.
	\newblock {\em Rev. Mod. Phys.}, 80:395--450, Apr 2008.
	
	\bibitem{Caffarel1994}
	Michel Caffarel and Werner Krauth.
	\newblock Exact diagonalization approach to correlated fermions in infinite
	dimensions: Mott transition and superconductivity.
	\newblock {\em Phys. Rev. Lett.}, 72:1545--1548, Mar 1994.
	
	\bibitem{Zgid2012}
	Dominika Zgid, Emanuel Gull, and Garnet Kin-Lic Chan.
	\newblock Truncated configuration interaction expansions as solvers for
	correlated quantum impurity models and dynamical mean-field theory.
	\newblock {\em Phys. Rev. B}, 86:165128, Oct 2012.
	
	\bibitem{Morin1992}
	M.~Morin, N.~J. Levinos, and A.~L. Harris.
	\newblock Vibrational energy transfer of co/cu(100): Nonadiabatic
	vibration/electron coupling.
	\newblock {\em The Journal of Chemical Physics}, 96(5):3950--3956, 1992.
	
	\bibitem{Wodtke2000}
	Yuhui Huang, Charles~T. Rettner, Daniel~J. Auerbach, and Alec~M. Wodtke.
	\newblock Vibrational promotion of electron transfer.
	\newblock {\em Science}, 290(5489):111--114, 2000.
	
	\bibitem{Baer1975}
	Michael Baer.
	\newblock Adiabatic and diabatic representations for atom-molecule collisions:
	Treatment of the collinear arrangement.
	\newblock {\em Chemical Physics Letters}, 35(1):112--118, 1975.
	
	\bibitem{Baer1980}
	Michael Baer.
	\newblock Electronic non-adiabatic transitions derivation of the general
	adiabatic-diabatic transformation matrix.
	\newblock {\em Molecular Physics}, 40(4):1011--1013, 1980.
	
	\bibitem{Pacher1988}
	T~Pacher, LS~Cederbaum, and H~K{\"o}ppel.
	\newblock Approximately diabatic states from block diagonalization of the
	electronic hamiltonian.
	\newblock {\em The Journal of chemical physics}, 89(12):7367--7381, 1988.
	
	\bibitem{Pacher1991}
	T.~Pacher, H.~K{\"o}ppel, and L.~S. Cederbaum.
	\newblock Quasidiabatic states from ab initio calculations by block
	diagonalization of the electronic hamiltonian: Use of frozen orbitals.
	\newblock {\em The Journal of Chemical Physics}, 95(9):6668--6680, 1991.
	
	\bibitem{Ruedenberg1993}
	Klaus Ruedenberg and Gregory~J Atchity.
	\newblock A quantum chemical determination of diabatic states.
	\newblock {\em The Journal of chemical physics}, 99(5):3799--3803, 1993.
	
	\bibitem{Ruedenberg1997}
	Gregory~J Atchity and Klaus Ruedenberg.
	\newblock Determination of diabatic states through enforcement of
	configurational uniformity.
	\newblock {\em Theoretical Chemistry Accounts: Theory, Computation, and
		Modeling (Theoretica Chimica Acta)}, 97(1):47--58, 1997.
	
	\bibitem{Truhlar2001}
	Hisao Nakamura and Donald~G Truhlar.
	\newblock The direct calculation of diabatic states based on configurational
	uniformity.
	\newblock {\em The Journal of Chemical Physics}, 115(22):10353--10372, 2001.
	
	\bibitem{Truhlar2002}
	Hisao Nakamura and Donald~G Truhlar.
	\newblock Direct diabatization of electronic states by the fourfold way. ii.
	dynamical correlation and rearrangement processes.
	\newblock {\em The Journal of chemical physics}, 117(12):5576--5593, 2002.
	
	\bibitem{Truhlar2003}
	Hisao Nakamura and Donald~G Truhlar.
	\newblock Extension of the fourfold way for calculation of global diabatic
	potential energy surfaces of complex, multiarrangement, non-born--oppenheimer
	systems: Application to hnco (s 0, s 1).
	\newblock {\em The Journal of chemical physics}, 118(15):6816--6829, 2003.
	
	\bibitem{Subotnik2008}
	Joseph~E Subotnik, Sina Yeganeh, Robert~J Cave, and Mark~A Ratner.
	\newblock Constructing diabatic states from adiabatic states: Extending
	generalized mulliken--hush to multiple charge centers with boys localization.
	\newblock {\em The Journal of chemical physics}, 129(24):244101, 2008.
	
	\bibitem{Subotnik2009}
	Joseph~E Subotnik, Robert~J Cave, Ryan~P Steele, and Neil Shenvi.
	\newblock The initial and final states of electron and energy transfer
	processes: Diabatization as motivated by system-solvent interactions.
	\newblock {\em The Journal of chemical physics}, 130(23):234102, 2009.
	
	\bibitem{Feshbach1962}
	Herman Feshbach.
	\newblock A unified theory of nuclear reactions. ii.
	\newblock {\em Annals of Physics}, 19(2):287 -- 313, 1962.
	
	\bibitem{OMalley1967}
	Thomas~F. O'Malley.
	\newblock Slow heavy-particle collision theory based on a quasiadiabatic
	representation of the electronic states of molecules.
	\newblock {\em Phys. Rev.}, 162:98--104, Oct 1967.
	
	\bibitem{Domcke1983}
	W.~Domcke.
	\newblock Projection-operator approach to potential scattering.
	\newblock {\em Phys. Rev. A}, 28:2777--2791, Nov 1983.
	
	\bibitem{Domcke1991}
	W.~Domcke.
	\newblock Theory of resonance and threshold effects in electron-molecule
	collisions: The projection-operator approach.
	\newblock {\em Physics Reports}, 208(2):97 -- 188, 1991.
	
	\bibitem{Kondov2007}
	Ivan Kondov, Martin {\v{C}}{\'\i}{\v{z}}ek, Claudia Benesch, Haobin Wang, and
	Michael Thoss.
	\newblock Quantum dynamics of photoinduced electron-transfer reactions in dye-
	semiconductor systems: First-principles description and application to
	coumarin 343- tio2.
	\newblock {\em The Journal of Physical Chemistry C}, 111(32):11970--11981,
	2007.
	
	\bibitem{Brandbyge2002}
	Mads Brandbyge, Jos\'e-Luis Mozos, Pablo Ordej\'on, Jeremy Taylor, and Kurt
	Stokbro.
	\newblock Density-functional method for nonequilibrium electron transport.
	\newblock {\em Phys. Rev. B}, 65:165401, Mar 2002.
	
	\bibitem{Levine2006}
	Benjamin~G. Levine, Chaehyuk Ko, Jason Quenneville, and Todd~J. Mart{\'I}nez.
	\newblock Conical intersections and double excitations in time-dependent
	density functional theory.
	\newblock {\em Molecular Physics}, 104(5-7):1039--1051, 2006.
	
	\bibitem{Gozem2014}
	Samer Gozem, Federico Melaccio, Alessio Valentini, Michael Filatov, Miquel
	Huix-Rotllant, Nicolas Ferr{\'e}, Luis~Manuel Frutos, Celestino Angeli,
	Anna~I. Krylov, Alexander~A. Granovsky, Roland Lindh, and Massimo Olivucci.
	\newblock Shape of multireference, equation-of-motion coupled-cluster, and
	density functional theory potential energy surfaces at a conical
	intersection.
	\newblock {\em Journal of Chemical Theory and Computation}, 10(8):3074--3084,
	2014.
	\newblock PMID: 26588278.
	
	\bibitem{Kaduk2012}
	Benjamin Kaduk.
	\newblock Constrained density functional theory.
	\newblock {\em Chemical reviews}, 112(1):321--370, 1 2012.
	
	\bibitem{Kaduk2010}
	Benjamin Kaduk and Troy Van~Voorhis.
	\newblock Communication: Conical intersections using constrained density
	functional theory--configuration interaction.
	\newblock {\em The Journal of Chemical Physics}, 133(6):061102, 2010.
	
	\bibitem{Grafenstein2005}
	J{\"u}rgen Gr{\"a}fenstein and Dieter~Cremer *.
	\newblock Development of a cas-dft method covering non-dynamical and dynamical
	electron correlation in a balanced way.
	\newblock {\em Molecular Physics}, 103(2-3):279--308, 2005.
	
	\bibitem{Grimme1999}
	Stefan Grimme and Mirko Waletzke.
	\newblock A combination of kohn--sham density functional theory and
	multi-reference configuration interaction methods.
	\newblock {\em The Journal of Chemical Physics}, 111(13):5645--5655, 1999.
	
	\bibitem{Maitra2004}
	Neepa~T. Maitra, Fan Zhang, Robert~J. Cave, and Kieron Burke.
	\newblock Double excitations within time-dependent density functional theory
	linear response.
	\newblock {\em The Journal of Chemical Physics}, 120(13):5932--5937, 2004.
	
	\bibitem{Cave2004}
	Robert~J. Cave, Fan Zhang, Neepa~T. Maitra, and Kieron Burke.
	\newblock A dressed tddft treatment of the 21ag states of butadiene and
	hexatriene.
	\newblock {\em Chemical Physics Letters}, 389(1):39 -- 42, 2004.
	
	\bibitem{Laikov2007}
	Dimitri Laikov and Spiridoula Matsika.
	\newblock Inclusion of second-order correlation effects for the ground and
	singly-excited states suitable for the study of conical intersections: The
	cis(2) model.
	\newblock {\em Chemical Physics Letters}, 448(1):132 -- 137, 2007.
	
	\bibitem{Teh2019}
	Hung-Hsuan Teh and Joseph~E. Subotnik.
	\newblock The simplest possible approach for simulating s0--s1 conical
	intersections with dft/tddft: Adding one doubly excited configuration.
	\newblock {\em The Journal of Physical Chemistry Letters}, 10(12):3426--3432,
	2019.
	\newblock PMID: 31135162.
	
	\bibitem{Knizia2012}
	Gerald Knizia and Garnet Kin-Lic Chan.
	\newblock Density matrix embedding: A simple alternative to dynamical
	mean-field theory.
	\newblock {\em Phys. Rev. Lett.}, 109:186404, Nov 2012.
	
	\bibitem{Wouters2016}
	Sebastian Wouters, Carlos~A. Jim{\'e}nez-Hoyos, Qiming Sun, and Garnet K.~L.
	Chan.
	\newblock A practical guide to density matrix embedding theory in quantum
	chemistry.
	\newblock {\em Journal of Chemical Theory and Computation}, 12(6):2706--2719,
	06 2016.
	
	\bibitem{Govind1998}
	N.~Govind, Y.A. Wang, A.J.R. da~Silva, and E.A. Carter.
	\newblock Accurate ab initio energetics of extended systems via explicit
	correlation embedded in a density functional environment.
	\newblock {\em Chemical Physics Letters}, 295(1):129 -- 134, 1998.
	
	\bibitem{Govind1999}
	Niranjan Govind, Yan~Alexander Wang, and Emily~A. Carter.
	\newblock Electronic-structure calculations by first-principles density-based
	embedding of explicitly correlated systems.
	\newblock {\em The Journal of Chemical Physics}, 110(16):7677--7688, 1999.
	
	\bibitem{Bruus2004}
	Henrik Bruus and Karsten Flensberg.
	\newblock {\em Many-body quantum theory in condensed matter physics: an
		introduction}.
	\newblock Oxford university press, 2004.
	
	\bibitem{Schonhammer1975}
	K.~Sch{\"o}nhammer.
	\newblock A variational method for hydrogen chemisorption.
	\newblock {\em Zeitschrift f{\"u}r Physik B Condensed Matter}, 21(4):389--392,
	Dec 1975.
	
	\bibitem{footnote:diab}
	Mathematically, this statement means that for any $M, [P_M, Q]\neq 0$.
	
	\bibitem{Lowdin1950}
	Per‐Olov L{\"o}wdin.
	\newblock On the non‐orthogonality problem connected with the use of atomic
	wave functions in the theory of molecules and crystals.
	\newblock {\em The Journal of Chemical Physics}, 18(3):365--375, 1950.
	
	\bibitem{Lowdin1970}
	Per-Olov L{\"o}wdin.
	\newblock On the nonorthogonality problem.
	\newblock In {\em Advances in quantum chemistry}, volume~5, pages 185--199.
	Elsevier, 1970.
	
	\bibitem{Carlson1957}
	B.~C. Carlson and Joseph~M. Keller.
	\newblock Orthogonalization procedures and the localization of wannier
	functions.
	\newblock {\em Phys. Rev.}, 105:102--103, Jan 1957.
	
	\bibitem{Fischer1965}
	Inga Fischer‐Hjalmars.
	\newblock Deduction of the zero differential overlap approximation from an
	orthogonal atomic orbital basis.
	\newblock {\em The Journal of Chemical Physics}, 42(6):1962--1972, 1965.
	
	\bibitem{Coulson1949}
	Prof.~C.A. Coulson and Miss~I. Fischer.
	\newblock Xxxiv. notes on the molecular orbital treatment of the hydrogen
	molecule.
	\newblock {\em The London, Edinburgh, and Dublin Philosophical Magazine and
		Journal of Science}, 40(303):386--393, 1949.
	
	\bibitem{Trail2003}
	J.~R. Trail, D.~M. Bird, M.~Persson, and S.~Holloway.
	\newblock Electron--hole pair creation by atoms incident on a metal surface.
	\newblock {\em The Journal of Chemical Physics}, 119(8):4539--4549, 2003.
	
	\bibitem{Tully1995}
	Martin Head‐Gordon and John~C. Tully.
	\newblock Molecular dynamics with electronic frictions.
	\newblock {\em The Journal of Chemical Physics}, 103(23):10137--10145, 1995.
	
	\bibitem{Dou2017prl}
	Wenjie Dou, Gaohan Miao, and Joseph~E. Subotnik.
	\newblock Born-oppenheimer dynamics, electronic friction, and the inclusion of
	electron-electron interactions.
	\newblock {\em Phys. Rev. Lett.}, 119:046001, Jul 2017.
	
	\bibitem{Jin2019}
	Zuxin Jin and Joseph~E. Subotnik.
	\newblock A practical ansatz for evaluating the electronic friction tensor
	accurately, efficiently, and in a nearly black-box format.
	\newblock {\em The Journal of Chemical Physics}, 150(16):164105, 2019.
	
	\bibitem{Tully1990}
	John~C. Tully.
	\newblock Molecular dynamics with electronic transitions.
	\newblock {\em The Journal of chemical physics}, 93(2):1061, 1990.
	
	\bibitem{Miao2017}
	Gaohan Miao, Wenjie Dou, and Joseph Subotnik.
	\newblock Vibrational relaxation at a metal surface: Electronic friction versus
	classical master equations.
	\newblock {\em The Journal of Chemical Physics}, 147(22):224105, 2017.
	
	\bibitem{Tully2009I}
	Neil Shenvi, Sharani Roy, and John~C. Tully.
	\newblock Nonadiabatic dynamics at metal surfaces: Independent-electron surface
	hopping.
	\newblock {\em The Journal of Chemical Physics}, 130(17), 2009.
	
	\bibitem{Tully2009M}
	Sharani Roy, Neil~A. Shenvi, and John~C. Tully.
	\newblock Model hamiltonian for the interaction of no with the au(111) surface.
	\newblock {\em The Journal of Chemical Physics}, 130(17):174716, 2009.
	
	\bibitem{Miao2019}
	Gaohan Miao, Wenjun Ouyang, and Joseph Subotnik.
	\newblock A comparison of surface hopping approaches for capturing
	metal-molecule electron transfer: A broadened classical master equation
	versus independent electron surface hopping.
	\newblock {\em The Journal of Chemical Physics}, 150(4):041711, 2019.
	
	\bibitem{Dou2016BCME}
	Wenjie Dou and Joseph~E. Subotnik.
	\newblock A broadened classical master equation approach for nonadiabatic
	dynamics at metal surfaces: Beyond the weak molecule-metal coupling limit.
	\newblock {\em The Journal of Chemical Physics}, 144(2):024116, 2016.
	
	\bibitem{Dou2017BCME}
	Wenjie Dou and Joseph~E. Subotnik.
	\newblock Electronic friction near metal surfaces: A case where molecule-metal
	couplings depend on nuclear coordinates.
	\newblock {\em The Journal of Chemical Physics}, 146(9):092304, 2017.
	
\end{thebibliography}
\end{document}